\documentstyle[12pt]{ioplppt} 
\newcommand{\be} {\begin{eqnarray}}
\newcommand{\ee} {\end{eqnarray}}
\newcommand{\beq} {\begin{equation}}
\newcommand{\eeq} {\end{equation}}
\newcommand{\bi} {\begin{itemize}}
\newcommand{\ei} {\end{itemize}}
\newcommand{\ben} {\begin{enumerate}}
\newcommand{\een} {\end{enumerate}}
\newcommand{\Det}[1]{{\rm Det}_{{#1}}}
\newcommand{\nno}{\nonumber}
\newcommand{\nn}{\nonumber}

\newcommand{\Ho}[2]{H^{(1)}_{#1} (k#2)}
\newcommand{\Ht}[2]{H^{(2)}_{#1} (k#2)}
\newcommand{\Jb}[2]{J_{#1} (k#2)}

\newcommand{\equa}[1]{(\ref{#1})}

\newcommand{\semiclass}{\ \stackrel{\rm s.c.}{\longrightarrow}\ }

\jl{1}
\begin{document}
\title{A direct link between the quantum-mechanical 
and semiclassical determination of scattering resonances}[A 
direct link between quantum-mechanical and semiclassical scattering]

\author{Andreas Wirzba\ftnote{2}
{Andreas.Wirzba@physik.tu-darmstadt.de}
and Michael Henseler\ftnote{3}{Present address: 
 Max-Planck-Institut f\"{u}r Physik
 komplexer Systeme, N\"othnitzer~Str.~38, \mbox{D-01187~Dresden}, 
Germany,
 michael@mpipks-dresden.mpg.de} }

\address{
 Institut f\"{u}r Kernphysik,
 Technische Universit\"at Darmstadt, 
 Schlo{\ss}gartenstra{\ss}e 9,\\
 D-64289 Darmstadt,
 Germany}
\begin{abstract}
We investigate the scattering of a point particle from $n$
non-overlapping, disconnected  hard disks which are fixed in the
two-dimensional plane and study the
connection between the spectral properties of the quantum-mechanical
scattering matrix and its semiclassical equivalent based on the
semiclassical zeta function of Gutzwiller and Voros.  
We rewrite the determinant of the
scattering matrix in such a way that it separates into the product
of  $n$ determinants of 1-disk scattering matrices -- 
representing
the incoherent part of the scattering from the $n$ disk 
system -- and
the ratio of two mutually complex conjugate determinants of the
genuine multi-scattering kernel, ${\bf M}$, which is of 
Korringa-Kohn-Rostoker-type and
represents the coherent multi-disk aspect of the $n$-disk scattering.
Our result is well-defined at every step of the calculation, as the
on-shell {\bf T}--matrix and the kernel ${\bf M}\mbox{$-$}{\bf 1}$
are shown to be trace-class.  We stress that the cumulant expansion
(which {\em defines} the determinant over an infinite, but 
trace class
matrix) induces the curvature regularization scheme to the
Gutzwiller-Voros zeta function and thus
leads to a new, well-defined and direct derivation of the
semiclassical spectral function.  We show that 
unitarity is preserved even at the semiclassical level.  
\end{abstract}
\maketitle

\section{Introduction\label{chap:intro}}
\setcounter{equation}{0}
In scattering problems whose classical
analog is completely hyperbolic or even chaotic, as e.g.\ $n$-disk
scattering systems,
the connection between the spectral properties of
exact quantum mechanics and
semiclassics  has been rather indirect
in the past. 
Mainly the resonance predictions of exact and semiclassical 
calculations have been compared which of course still is
a useful exercise, but does not fully capture the rich structure
of the problem. 
As shown in ref.~\cite{cvw96}, there exist several semiclassical 
spectral functions
which predict the very same leading resonances but give different
results for the phase shifts. 
Similar results
are known for bound systems, see refs.~\cite{KeatSieb94,Keat97}: 
the comparison of the
analytic structure of the pertinent spectral determinant with various
semiclassical zeta functions furnishes the possibility of 
making much more
discriminating tests of the 
semiclassical approximation than the mere 
comparison of exact eigenvalues 
with the corresponding semiclassical predictions. 

In the exact
quantum-mechanical calculations
the resonance poles are extracted from 
the zeros of a characteristic scattering
determinant (see e.g.\ \cite{gr}), 
whereas the semiclassical predictions follow from the zeros
(poles)
of a semiclassical spectral determinant (trace) 
of Gutzwiller~\cite{gutzwiller} 
and Voros~\cite{voros88}. These semiclassical quantities have 
either been {\em formally} (i.e.\ without induced 
regularization prescription)
taken over from bounded problems (where  the semiclassical reduction
is done via the spectral
density)~\cite{scherer,pinball} or they have  
been extrapolated from the
corresponding {\em classical} 
scattering determinant~\cite{fredh,fredh2}.
Here, our aim is to construct a {\em direct} link between the
quantum-mechanical and 
and the semiclassical treatment of hyperbolic scattering
in a concrete context, the $n$-disk repellers. 
The latter 
belong to the simplest realizations of hyperbolic 
or even chaotic scattering 
problems, since they have the structure of a 
quantum billiard --- without any
confining (outer) walls. Special emphasis is given to a well-defined 
quantum-mechanical starting-point which 
allows for the semiclassical reduction {\em including} 
the appropriate regularization
prescription. In this context the word ``direct'' 
refers to a link which is
not of {\em formal} nature, but includes 
a proper regularization prescription
which is {\em inherited} from quantum mechanics, 
and not {\em imposed from
the outside by hand}.

The $n$-disk problem   consists in 
the scattering of a
scalar point particle from $n>1$ circular, non-overlapping, 
disconnected hard disks which are fixed  in the
two-dimensional plane.  
Following the methods of Gaspard and Rice~\cite{gr} we construct
the pertinent 
on-shell {\bf T}--matrix which 
splits into the product of three matrices
${\bf C}(k) {\bf M}^{-1}(k) {\bf D}(k)$. The matrices
${\bf C}(k)$ and ${\bf D}(k)$ couple the incoming and
outgoing scattering
wave   (of wave number $k$), respectively, to {\em one} of the disks, 
whereas the matrix ${\bf M}(k)$ parametrizes
the scattering interior, i.e., the  {\em multi-scattering}
evolution in the
multi-disk geometry.
The  understanding is that
the resonance poles of the $n>1$ disk problem 
can only result from the zeros of the
characteristic  
determinant $\det{ }{\bf M}(k)$; 
see the quantum mechanical construction of
Gaspard and Rice~\cite{gr} for the three-disk
scattering system~\cite{Eck_org,gr_cl,gr_sc,Cvi_Eck_89}.
Their work relates to Berry's 
application~\cite{Berry_KKR,Berry_LH} of the
Korringa-Kohn-Rostoker (KKR) method~\cite{KKR}
to the (infinite) 
two-dimensional
Sinai-billiard problem which in turn is
based on Lloyd's multiple scattering 
method~\cite{Lloyd,Lloyd_smith}
for a finite cluster of  
non-overlapping muffin-tin potentials in three dimension.

On the semiclassical side, the geometrical
primitive periodic orbits (labelled by $p$) 
are       summed up -- including repeats (labelled by $r$) -- in the
Gutzwiller-Voros zeta function~\cite{gutzwiller,voros88,fredh} 
\be
  Z_{GV} (z;k) &=& 
 \exp\left\{ - \sum_{p} \sum_{r=1}^{\infty} \frac{1}{r}
 \frac{\left( z^{n_p}\, t_p(k) \right)^r }
 {1-\frac{1}{\Lambda_p^r}} \right\} 
 \label{GV_zeta_formal} \\
 &=& \prod_{p} \prod_{j=0}^{\infty}\left (1 - \frac{z^{n_p} t_p(k)}
{{\Lambda_p}^j}\right) \label{GV_zeta_formal_prod} \; ,
\ee 
where
$t_p(k)={\rm e}^{{\rm i} k L_p - {\rm i} \nu_p \pi/2}/
\sqrt{|\Lambda_p|}$ is the so-called
$p^{\,\rm th}$ cycle, $n_p$ is its topological length and $z$ is a 
book-keeping variable for keeping track of the topological order.   
The input is purely geometrical, i.e., the
lengths $L_p$, the Maslov indices $\nu_p$, and the stabilities (the
leading eigenvalues of the stability matrices) $\Lambda_p$ of the
$p^{\,\rm th}$ primitive periodic orbits. Note that both expressions 
for the Gutzwiller-Voros zeta function, the original one 
\equa{GV_zeta_formal} and the reformulation in 
terms of an infinite product
\equa{GV_zeta_formal_prod}, 
are purely formal. In general, they may not  
exist without regularization. (An exception is 
the non-chaotic 2-disk system, since it has only one periodic orbit, 
$t_0(k)$.)\
Therefore,
the semiclassical resonance
poles are normally computed from $Z_{GV}(\mbox{$z$=1};k)$ in the 
(by hand imposed) curvature
expansion~\cite{fredh,pinball,artuso} up to a given topological length
$m$.
This procedure corresponds to a Taylor expansion of $Z_{GV}(z;k)$
in $z$ around $z=0$ up to order $z^m$ 
(with $z$ taken to be one at the end): 
\be
 \fl
  Z_{GV} (z;k) = z^0 - z\sum_{n_p=1} \frac{t_p}{1-\frac{1}{\Lambda_p}}
 \\ 
  \lo- \frac{z^2}{2}\left\{ \sum_{n_p=2}\frac{2
 t_p}{1\!-\!\frac{1}{\Lambda_p}} \mbox{}+ \sum_{n_p=1}
 \frac{(t_p)^2}{1\!-\!\left (\frac{1}{\Lambda_p}\right )^2} -
 \sum_{n_p=1}\sum_{n_{p'}=1} \frac{t_p}{1\!-\!\frac{1}{\Lambda_p}}
 \frac{t_{p'}}{1\!-\!\frac{1}{\Lambda_{p'}}}\right\} +\cdots\; . 
 \label{gutzcurv} 
\ee 
This is one way of regularizing the formal expression of
the Gutzwiller-Voros zeta function
\equa{GV_zeta_formal}. The hope is that the 
limit $m\to\infty$ exists ---
at least in the semiclassical regime 
${\rm Re}\, k \gg 1/a$ where $a$ is
the characteristic length of the scattering potential.
We will show below that in the quantum-mechanical analog --- 
the cumulant expansion -- this limit can be taken.

As mentioned, the connection between quantum
mechanics and semiclassics for these scattering problems has been
the comparison of the corresponding resonance poles, the
zeros of the characteristic determinant on the one hand and the zeros
of the Gutzwiller-Voros zeta function -- in general in the curvature 
expansion -- on the other hand. In the literature (see e.g.\
\cite{scherer,pinball,gr_sc} based on \cite{bb_1,thirring})
this link is motivated by the 
semiclassical limit of the left hand sides 
of the
Krein-Friedel-Lloyd sum for the 
[integrated] spectral density~\cite{Krein,Friedel} 
and \cite{Lloyd,Lloyd_smith}
\be
 \lim_{\epsilon\to +0}\lim_{b\to \infty} \left ( 
 N^{(n)}(k+\i\epsilon;b)-N^{(0)}(k+\i\epsilon;b) \right )
 &=&\frac{1}{2\pi} {\rm Im} \Tr{ }\ln {\bf S}(k)\ ,
 \label{friedel_sum} \\
 \lim_{\epsilon\to +0}\lim_{b\to \infty} \left ( 
 \rho^{(n)}(k+\i\epsilon;b)- \rho^{(0)}(k+\i\epsilon;b) \right )
 &=& \frac{1}{2\pi} {\rm Im} 
 \Tr{ }\frac{d}{dk} \ln {\bf S}(k) \label{friedel_sumr}
 \; .
\ee 
See also \cite{faulkner}
for a modern discussion of the Krein-Friedel-Lloyd formula and 
\cite{thirring,gasp_varena} for the connection of the 
\equa{friedel_sumr} to the Wigner time-delay. 
In this way the scattering
problem is replaced by the difference of two bounded 
circular reference billiards
of the same radius $b$ which eventually will 
be taken to infinity, where
one contains in its interior the scattering configuration and the
other one is empty.
Here,
$\rho^{(n)}(k;b)$ [$N^{(n)}(k;b)$] and $\rho^{(0)}(k;b)$ 
[$N^{(0)}(k;b)$] 
are the 
spectral densities [integrated spectral densities] in the
presence or absence of the scatterers, respectively.
In the semiclassical limit, they will be 
replaced by a smooth Weyl term and
an oscillating periodic orbit sum.
Note that the above expressions 
make only sense for wave numbers $k$ above
the real axis. Especially, if $k$ is 
chosen to be real, $\epsilon$ must be
greater than zero. Otherwise, 
the exact left-hand sides would give discontinuous
staircase or delta functions, respectively, 
whereas the right-hand sides are
by definition continuous functions of $k$. Thus, the order of the
two limits in \equa{friedel_sum} 
and \equa{friedel_sumr} is important,
see, e.g., Balian and Bloch ~\cite{bb_1} 
who stress that smoothed level
densities should be inserted into the Friedel sums. 

We stress that these links are of {\em indirect} nature, 
since unregulated 
expressions for the semiclassical
Gutzwiller trace formula for {\em bound} systems arise on the 
left-hand sides of the (integrated) Krein-Friedel-Lloyd sums in the
semiclassical reduction. Neither the
curvature regularization scheme 
nor other constraints on the periodic orbit
sum follow from this in a natural way. Since the indirect link of 
\equa{friedel_sum} and \equa{friedel_sumr}
is made with
the help of bound systems, the question might arise for instance 
whether in scattering systems 
the Gutzwiller-Voros zeta function 
should be resummed according to Berry and
Keating~\cite{berry_keats} or 
not.
This question is answered by the presence 
of the $\i\epsilon$ term {\em and}
the second limit. The wave number 
is shifted by this from the real axis
into the upper complex $k$ plane. 
This corresponds to
a ``de-hermitezation'' of the underlying hamiltonian -- the  
Berry-Keating resummation which explicitly 
makes use of the reality 
of the eigen-energies of a {\em bound-system}
does not apply here.
The necessity of the $+\i\epsilon$ in the semiclassical calculation
can be understood by purely phenomenological considerations:
Without the $+\i\epsilon$ term 
there is no reason why one should be able 
to neglect spurious periodic orbits which solely exist
because of the introduction of the confining boundary. 
The subtraction of the 
second (empty) reference system helps just in 
the removal of those spurious 
periodic orbits which never encounter the scattering region. 
The ones 
that do so would still survive the first limit $b \to \infty$, 
if they
were not damped out by the $+\i\epsilon$ term.

The expression for the integrated spectral densities 
is further complicated
by the fact that the $\epsilon$-limit and the 
integration do not commute
either. As a consequence there appears on the l.h.s.\  
of \equa{friedel_sum} 
an (in general)
undetermined integration constant.

Independently of this comparison via the Krein-Friedel-Lloyd sums,
it was shown in \cite{aw_chaos} that the
characteristic determinant $\det{ } {\bf M}(k) =\det{ }({\bf
1}+{\bf A}(k))$ can be re-arranged via ${\rm e}^{ \Tr{ } \ln( {\bf
1}+{\bf A}(k))}$ in a cumulant expansion and that the semiclassical
analogs to the first traces, $\Tr{ }({\bf A}^m (k))$
($m=1,2,3,\dots$), contain (including creeping periodic orbits) the
sums of all periodic orbits (with and without repeats) of total
topological length $m$. Thus \equa{gutzcurv} should be directly
compared with its quantum analog, the cumulant expansion  
\beq
\fl 
  \det{ }({\bf 1}+z {\bf A}) = 1 - (-z)
 \Tr{ } [{\bf A}(k)] -\frac{z^2}{2} \left \{    \Tr{ }[{\bf
 A}^2(k)] - \left [ \Tr{ } {\bf A}(k) \right ]^2 \right \} +
 \cdots \ .  
 \label{standard_cum}
\eeq 
The knowledge of the traces is sufficient to organize
the cumulant expansion of the determinant
\be
 \det{ }({\bf 1}+z {\bf A})= \sum_{m=0}^{\infty}\, z^m c_m({\bf A})
 \label{cumulant-sum}
\ee
(with $c_0({\bf A})\equiv 1$)
in terms of a recursion relation for the cumulants
(see the discussion of the Plemelj-Smithies formula in
in the appendix) 
\be
 c_m({\bf A}) &=& \frac{1}{m}\sum_{k=1}^{m} 
  (-1)^{k+1} c_{m-k}({\bf A})\, 
   \Tr{ }({\bf A}^k) \qquad {\rm for}\ m \geq 1 \; . 
 \label{cumulant-recursion}  
\ee

In the 2nd paper of \cite{aw_chaos} the geometrical semiclassical
analogs to the first three traces were explicitly constructed for the
2-disk problem.  The so-constructed geometrical terms correspond
exactly (including all prefactors, Maslov indices, 
and symmetry reductions) 
to the once, twice or
three-times repeated periodic orbit that is  spanned by the two disks.
(Note that the two-disk system 
has only one classical periodic orbit.)\   
In the mean-time,
one of us has shown that, with the help of 
Watson resummation techniques~\cite{Watson,franz} and 
by complete induction, 
the semiclassical reduction of the quantum mechanical traces of any
non-overlapping $2\leq n <\infty$ disk system [where in addition  
grazing or penumbra orbits~\cite{Nussenzveig,penumbra} have to be 
avoided in order
to guarantee unique isolated saddle point contributions] 
reads as follows~\cite{aw_unpub}
\beq
\fl
 (-1)^m \Tr{ } ({\bf A}^m(k) ) \semiclass 
 \sum_{p}\sum_{r>0}\, 
\delta_{m,r n_p}\,
n_p\, 
\frac{{t_p(k)}^r}{{1\!-\!\left
 (\frac{1}{\Lambda_p} \right )^r}} 
   \ + \ {\rm diffractive\, creeping\,
 orbits}, 
 \label{trace-semiclass}
\eeq 
where $t_p$ are periodic orbits of topological 
length $n_p$ with $r$ repeats. 
The semiclassical reduction \equa{trace-semiclass} holds of 
course only in the case that 
${\rm Re}\, k$ is big enough compared with the inverse of
the smallest length scale. Note that \equa{trace-semiclass} 
does not imply that the semiclassical limit 
$k\to \infty$ and the cumulant limit 
$m\to\infty$ commute in general, i.e.,
that the curvature expansion exists.
The factor $n_p$ results from the count of the cyclic 
permutations of a ``symbolic word'' of length $n_p$ 
which all label the same primary periodic
orbit  $t_p$. As the leading 
semiclassical approximation to $\Tr{ } ({\bf A}^m(k) )$ is based 
on the replacement of the $m$ sums by $m$ integrals which are then 
evaluated according to the 
saddle point approximation, the qualitative 
structure of the r.h.s.\ of 
\equa{trace-semiclass} is expected. 
The nontrivial points are the weights, the
phases, and the pruning of ghost orbits 
which according to \cite{aw_unpub} 
follows the scheme presented  in 
\cite{Berry_KKR}.
In \cite{alonso,gasp_hbar,vattay_hbar} $\hbar$-corrections 
to the geometrical
periodic orbits were constructed, whereas the authors of
\cite{vwr_prl} extended the Gutzwiller-Voros zeta function to
include diffractive creeping periodic orbits as well. 

By inserting the semiclassical approximation \equa{trace-semiclass} of
the traces into the exact 
recursion relation \equa{cumulant-recursion},
one can
find  a compact expression
of the curvature-regularized version of the  
Gutzwiller-Voros zeta function~\cite{fredh,pinball,artuso}:
\be
 Z_{GV}(z;k)=\sum_{m=0}^\infty\, z^m  c_m({\rm s.c.}) \; ,
 \label{curvature-sum}
\ee
(with $c_0({\rm s.c.})\equiv 1$),
where the curvature terms $ c_m({\rm s.c.})$ satisfy 
the semiclassical recursion relation
\beq
\fl
 c_m({\rm s.c.}) = -\frac{1}{m}\sum_{k=1}^{m} 
 c_{m-k}({\rm s.c.})
 \sum_{p}\sum_{r>0} \,\delta_{k,r n_p}\, n_p\, 
\frac{{t_p(k)}^r}{{1-\left
 (\frac{1}{\Lambda_p} \right )^r}}  \qquad {\rm for}\ m \geq 1 \; . 
\label{curvature-recursion} 
\eeq

Below, we construct explicitly a {\em direct} link between  the  full 
quantum-mechanical
 {\bf S}--matrix and the Gutzwiller-Voros zeta function
in the particular case of $n$-disk scattering. 
We will show that 
{\em all} necessary steps in the 
quantum-mechanical description are justified.
It is demonstrated that the spectral determinant of 
the $n$-disk problem  splits uniquely
into a product of $n$ incoherent
one-disk terms and one coherent genuine
multi-disk term which under 
suitable symmetries separates into distinct
symmetry classes. Thus, we have found 
a well-defined starting-point for the
semiclassical reduction.   
Since  the {\bf T}--matrix and 
the matrix
${\bf A}\equiv {\bf M}-{\bf 1}$ are trace class matrices 
(i.e., the
sum of the diagonal matrix elements is 
absolutely converging in any orthonormal basis), 
the corresponding determinants of the $n$-disk 
and one-disk {\bf S}-matrices 
and the
characteristic matrix {\bf M} are guaranteed to exist 
although they are infinite matrices. 
The cumulant expansion defines 
the characteristic determinant and 
guarantees a finite, unambiguous result.
As the semiclassical limit is taken, 
the defining quantum-mechanical cumulant 
expansion reduces  to the curvature-expansion-regularization 
of the semiclassical
spectral function.
It will also be shown
that unitarity is preserved at the semiclassical
level under the precondition that the 
curvature sum converges or is suitably
truncated.
In Appendix~A
the trace-class properties of all matrices entering the expression
for the $n$-disk {\bf S}--matrix will  be shown explicitly. 
  
\section{Direct link\label{chap:link}}
\setcounter{equation}{0}
If one is only interested in spectral 
properties (i.e., in resonances
and not in wave functions)
it is sufficient to construct the determinant,
$\det{ } {\bf S}$,  of the 
scattering matrix ${\bf
S}$.   The determinant  
is invariant under any change of a
complete basis representing 
the $\bf S$-matrix. (The determinant of ${\bf S}$ is
therefore also independent of the coordinate system.)

For any non-overlapping system of $n$-disks (which may even have
different sizes, i.e., different disk-radii: $a_j$, $j=1,\dots,n$)
the $\bf S$-matrix can be split up in the following way \cite{mh}
using the methods and notation of Gaspard and Rice \cite{gr}  (see
also \cite{Lloyd_smith}): 
\be 
 {\bf S}_{m m'}^{(n)}(k) 
 = \delta_{m m'} - {\rm i\,} 
 {\bf C}_{m l}^{\ \ j}(k) \,{ \left \{ {\bf M}^{-1}(k)\right
 \} }_{l l'}^{j j'}\, {\bf D}_{l'm'}^{j'}(k) \; ,
\label{Smatrix} 
\ee 
where
$j,\,j'=1,\dots,n$ (with $n$ finite) 
label the ($n$) different disks and
the quantum numbers $-\infty < m,m',l,l' < +\infty $ refer to a
complete set of spherical eigenfunctions, $\{|m\rangle \}$, with
respect to the origin of 
the 2-dimensional plane (repeated indices are
of course summed over). 
The matrices $\bf  C$ and $\bf D$ can be found in 
Gaspard and Rice \cite{gr}; 
they
depend on the origin and orientation of the global coordinate system
of the two-dimensional plane and are separable in the disk index $j$.
They 
parameterize  the coupling of the incoming and outgoing
scattering wave, 
respectively, to the $j^{\, \rm th}$ disk  and describe
therefore the single-disk aspects 
of the scattering of a point particle from the
$n$ disks:
\be
 {\bf C}_{m l}^{\ \ j} &=&\frac{2{\rm i}}{\pi a_j}
    \frac{ J_{m-l}(k R_j)}{\Ho{l}{a_j}} 
   {\rm e}^{{\rm i} m \Phi_{R_j}}\; , \label{Cmatrix} \\
 {\bf D}_{l'm'}^{j'} &=&
   - \pi a_{j'} J_{m'-l'}(k R_{j'}) J_{l'} (ka_{j'}) 
     {\rm e}^{-{\rm i} m' \Phi_{R_{j'}}}\; . \label{Dmatrix} 
\ee
Here $R_j$  and $\Phi_{R_j}$ denote 
the distance  and angle, respectively, of the
ray from the origin in the 2-dimensional plane to
the center of the disk $j$ as measured in 
the global coordinate system. 
$\Ho{l}{r}$  is the ordinary Hankel function of first kind and 
$\Jb{l}{r}$ the corresponding ordinary Bessel function. 
The matrix $\bf M$ is the genuine multi-disk ``scattering'' 
matrix with eliminated single-disk
properties  (in the pure
1-disk scattering case $\bf M$ becomes just the identity 
matrix)~\cite{mh}:
\be                                         
    {\bf M}_{l l'}^{j j'} = \delta_{jj'} \delta_{l l'} +
     (1-\delta_{jj'})\, \frac{a_j}{a_{j'}}\, 
           \frac{ \Jb{l}{a_j} }{\Ho{l'}{a_{j'}} }\,
                      \Ho{l-l'}{R_{jj'}} \,
                        \Gamma_{jj'}(l,l') \; .
 \label{Mmatrix} 
\ee
It has the structure of a KKR-matrix (see 
\cite{Berry_KKR,Berry_LH,Lloyd_smith})
and is the generalization of the 
result of Gaspard and Rice~\cite{gr} 
for the equilateral 3-disk system to
a general $n$-disk configuration where the disks 
can have {\em different} 
sizes.
Here, $R_{jj'}$ is the separation between the centers of
the $j$th and $j'$th disk and $R_{jj'} = R_{j'j}$. The matrix
$\Gamma_{jj'}(l,l')=   {\rm e}^{{\rm i}( l\alpha_{j' j}
          -l'(\alpha_{j j'} -\pi))}$ contains -- besides a
phase factor --
the angle 
$\alpha_{j'j}$ of the ray from the center of disk $j$ to
the center of disk $j'$ 
as measured in the local (body-fixed) coordinate
system of disk $j$.
Note that $ \Gamma_{jj'}(l,l')  
= (-1)^{l-l'} (\, \Gamma_{j'j}(l',l) \,)^{\ast}$.
The Gaspard and Rice prefactors, 
i.e., $(\pi a / 2 i)$, of $\bf M$ are rescaled 
into $\bf C$ and $\bf D$.
The product ${\bf C} {\bf M}^{-1} {\bf D}$ corresponds to the
three-dimensional result of Lloyd and Smith 
for the on-shell ${\bf T}$-matrix of a 
finite cluster of
non-overlapping muffin-tin potentials. 
The expressions of Lloyd and
Smith (see
(98) of \cite{Lloyd_smith} and also Berry's 
form~\cite{Berry_KKR}) at first sight seem
to look simpler than ours and the ones of \cite{gr} for the
3-disk system, 
as, e.g., in ${\bf M}$ the asymmetric term
$ a_j \Jb{l}{a_j}/ a_{j'} \Ho{l'}{a_{j'}}$ is replaced by a symmetric 
combination, $\Jb{l}{a_j}/\Ho{l}{a_j}$.
This form, however, is not of trace-class. Thus, manipulations 
which are
allowed within our description are not necessarily allowed in Berry's
and Lloyd's formulation.
After a {\em formal} rearrangement of our matrices
we can derive the result of Berry and Lloyd.
Note, however, that the trace-class property of ${\bf M}$ is
lost in this formal manipulation, 
such that the infinite determinant and the corresponding cumulant 
expansion converge only conditionally, and not absolutely as in our  
case.

The $l$-labelled matrices ${\bf S}^{(n)}-{\bf 1}$, ${\bf C}$ and
${\bf D}$ as well 
as the $ \{l,j\}$-labelled  matrix ${\bf M}-{\bf 1}$
are of ``trace-class''
(see the appendix for the proofs).
A matrix is called ``trace-class'', 
if, independently of the choice of the
orthonormal basis, the
sum of the diagonal matrix elements 
converges absolutely; it is called ``Hilbert-Schmidt'', if the sum of
the absolute squared diagonal matrix elements converges,   
see the appendix and M.~Reed and B.~Simon, Vol.1 and 4 
\cite{rs1,rs4} 
for the definitions and properties of trace-class and Hilbert-Schmidt 
matrices. Here, we will list only the most important ones:
(i) any trace-class matrix can be represented as the product of two
Hilbert-Schmidt matrices and any such product is trace-class; 
(ii)
the linear combination of a finite number 
of trace-class matrices is again
trace-class; (iii) the hermitean-conjugate 
of a trace-class matrix is again trace-class;
(iv) the product of two Hilbert-Schmidt
matrices or of a trace-class and a bounded matrix is trace-class and
commutes under the trace; (v) if ${\bf B}$ is trace-class, 
the determinant
$\det{}({\bf 1}+z {\bf B})$ exists and 
is an entire function of $z$; (vi) the
determinant is invariant under unitary transformations.
Therefore for all fixed values of $k$ (except at 
$k \leq 0$  
[the branch cut of the Hankel functions] and the countable
isolated zeros of $\Ho{m}{a_j}$ and  of $\Det{ } {\bf M}(k)$) 
the following operations
are mathematically allowed:
\be
     \det{ } {\bf S}^{(n)} &=&  \det{ }  \left( {\bf 1} 
    - {\rm i\,} {\bf C} {\bf M}^{-1} {\bf D} \right )
          = \exp\tr{ } \ln \left( {\bf 1} 
    - {\rm i\,} {\bf C} {\bf M}^{-1} {\bf D} \right )
     \nno \\
           &=& \exp \left\{- \sum_{N=1}^\infty \frac{{\rm i}^N}{N} 
      \tr{ } \left[ 
            \left( {\bf C} {\bf M}^{-1} {\bf D} \right)^N \right]  
              \right \} 
     \nno \\         
           &=& \exp \left\{-  \sum_{N=1}^\infty \frac{{\rm i}^N}{N} 
    \Tr{ } \left[  
    \left ({\bf M}^{-1} {\bf D} {\bf C} \right )^N \right ]  \right\} 
   \nno \\
            &=& \exp \Tr{ } 
   \ln\left ({\bf 1}-{\rm i\,} {\bf M}^{-1} {\bf D} {\bf C} \right )    
           = \Det{ } \left({\bf 1} -{\rm i\,} 
                 {\bf M}^{-1} {\bf D} {\bf C} \right )
           \nno \\                                     
          &=& \Det{ } \left [{\bf M}^{-1} ({\bf M} -{\rm i\,} 
        {\bf D C})\right]
    \nno \\
          &=& \frac{ \Det{ } ({\bf M} 
              - {\rm i\,} {\bf DC})} { \Det{ } ({\bf M}) } \; .
     \label{recoupl}
\ee
Actually,
$\det{}({\bf 1}+\mu{\bf A}) 
         = \exp\{- \sum_{N=1}^{\infty} \frac{(-\mu)^N}{N}
 \rm{tr}[ {\bf A}^N]\}$ 
is only valid for $|\mu \, {\rm max}\, \lambda_i| < 1$ 
where $\lambda_i$
is the $i$-th eigenvalue of ${\bf A}$. The determinant is directly
defined through its cumulant expansion 
(see equation (188) of \cite{rs4}) which
is therefore the analytical continuation 
of the ${\rm e}^{\tr{ } \log}$
representation. Thus the  ${\rm e}^{\tr{ } \log}$ notation 
should here be understood
as a compact
abbreviation for the defining cumulant expansion.  
The capital index
$L$ is a  multi-index $L=(l,j)$.
On the l.h.s.\ of \equa{recoupl} 
the determinant and traces are only taken over small $l$,
on the r.h.s.\ they are taken over multi-indices $L=(l,j)$ 
(we will use
the following convention: $\det{}\dots$ and $\tr{}\dots$ 
refer to the 
$|m\rangle$ space, $\Det{}\dots$ and $\Tr{}\dots$ refer 
to the multi-spaces). 
The corresponding  complete basis is now 
$\{ |L\rangle\}  = \{|m;j\rangle \}$
which now refers to the origin of the $j$th disk 
(for fixed $j$ of course)
and not to the origin of the 
2-dimensional plane any longer.
In deriving \equa{recoupl} the following facts have been used: 
\begin{description}
\item[(a)\ ]      
${\bf D}^j, {\bf C}^j$  -- if their index $j$ is kept fixed -- 
are of trace class (see the appendix)
\item[(b)\ ]
and therefore the product
${\bf D C}$  -- now in the multi-space $\{ |L\rangle \}$ --  
is of trace-class as long as $n$ is finite 
(see property (ii)), 
\item[(c)\ ]
${\bf M}-{\bf 1}$ is of trace-class (see the appendix). Thus
the determinant 
$\Det{ }\ {\bf M}(k)$ exists.
\item[(d)\ ]
${\bf M}$ is bounded, since 
it is the sum of a bounded and a trace-class matrix.
\item[(e)\ ]
${\bf M}$ is invertible everywhere where 
$\Det{ } {\bf M}(k)$ is defined 
(which excludes a countable number of zeros of the Hankel 
functions $\Ho{m}{a_j}$ and the
negative real $k$ axis, since there is a branch cut) 
and nonzero (which excludes a countable number
of isolated points in the lower 
$k$-plane) -- see the appendix for these properties.
Therefore and because of {\bf (d)} the matrix
${\bf M}^{-1}$ is bounded.
\item[(f)\ ]
${\bf C} {\bf M}^{-1} {\bf D}$, ${\bf M}^{-1} {\bf D C}$ 
are all of trace-class, since they are the product 
of bounded times 
trace-class matrices, and 
$\tr{ }[({\bf C} {\bf M}^{-1} {\bf D})^N]=
\Tr{ }[( {\bf M}^{-1} {\bf D C})^N ]$, because such products have
the cyclic permutation property under the trace  
(see properties (ii)  and
(iv)).
\item[(g)\ ]
 ${\bf M}-{\rm i\,}{\bf DC}-{\bf 1}$ is of  
trace-class because of the rule 
that the sum of two trace-class matrices is
again trace-class (see property (ii)).
\end{description}
Thus all the traces and determinants appearing in \equa{recoupl} 
are well-defined, except at the above mentioned $k$ values. 
Note that in the $\{|m;j\rangle\}$ basis the trace of 
${\bf M}-{\bf 1}$ vanishes trivially because of the 
$\delta_{jj'}$ terms in
\equa{Mmatrix}. This does not prove the trace-class property of 
${\bf M}-{\bf 1}$, since the finiteness (here vanishing) of 
$\Tr{ } ({\bf M}-{\bf 1} )$ has to be shown 
for every complete orthonormal
basis. After symmetry reduction (see below) 
$\Tr{ } ({\bf M}-{\bf 1} )$, calculated for any irreducible
representation, does not vanish any longer. However, the sum of the
traces of all irreducible representations weighted by their pertinent
degeneracies still vanishes of course. 
Semiclassically, this corresponds
to the fact that only in the fundamental domain there can exist 
one-letter ``symbolic words''.

Now, the computation of the determinant of the {\bf S}--matrix 
is  very much
simplified
in comparison with the original formulation,
since the last term of \equa{recoupl} is completely written 
in terms of closed form
expressions and does not involve ${\bf M}^{-1}$ any
longer.
Furthermore, using the notation of Gaspard and Rice \cite{gr}, 
one can easily
construct 
\be 
\fl {\bf M}_{l l'}^{j j'} 
    -{\rm i\,} {\bf D}_{l m'}^{j} {\bf C}_{m' l'}^{\ \ \ j'} 
 = \delta_{jj'} \delta_{ll'} \left (-
 \frac{\Ht{l'}{a_{j'}}} {\Ho{l'}{a_{j'}}} \right ) \nno \\ 
 -(1-\delta_{jj'})\, \frac{a_j}{a_{j'}}\,
 \frac{\Jb{l}{a_j}}{\Ho{l'}{a_{j'}}} \, \Ht{l-l'}{R_{jj'}} \,
 \Gamma_{jj'}(l,l') \; ,
\label{rewri} 
\ee 
where $\Ht{m}{r}$ is the
Hankel function of second kind.  Note that 
$ \{H^{(2)}_{m} (z)\}^\ast =H^{(1)}_{m}(z^\ast)$.  
The scattering from a single disk is a separable problem
and the  $\bf
S$-matrix for the 1-disk problem with the center at the origin reads 
\be
 {\bf S}^{(1)}_{ll'} (ka) = -\frac{\Ht{l}{a}} {\Ho{l}{a}} \, 
 \delta_{ll'} \; .\label{S1disk} 
\ee 
This can be seen by comparison of the general asymptotic expression
for the wavefunction with the exact solution for the 1-disk
problem~\cite{mh}.
Using \equa{rewri} and \equa{S1disk}
and trace-class properties of 
${\bf M}-{\bf 1}$, ${\bf M} - {\rm i\,} {\bf DC}-{\bf 1}$ 
and ${\bf S}^{(1)} -{\bf 1}$  
one can easily rewrite the
r.h.s.\ of \equa{recoupl} as
\beq 
\fl
 \det{ } {\bf S}^{(n)}(k) 
 = \frac{ \Det{ } ({\bf M}(k) - {\rm i\,} {\bf
 D}(k){\bf C}(k)) } {\Det{ } {\bf M}(k)}  
 = \left \{
 \prod_{j=1}^{n} \left( \det{ } {\bf S}^{(1)}(k a_j) \right)
 \right \}\, \frac{\Det{ }{ {\bf M}(k^\ast) }^\dagger}
{\Det{ } {\bf  M}(k)} \label{qm}\; ,
\eeq 
where now the zeros of the 
Hankel functions $H^{(2)}_m (ka_j)$ have to
be excluded as well.
In general, the single disks
have different sizes. 
Therefore they are labelled by the index $j$. Note
that the analogous formula for the three-dimensional scattering 
of a point particle from $n$
non-overlapping  balls (of different sizes in general) 
is structurally
completely the same~\cite{mh,N-ball} 
(except that the negative $k$-axis
is not excluded since  the spherical Hankel
functions have no branch cut).
In the above calculation it was used that
$\Gamma_{jj'}^\ast(l,l') = \Gamma_{jj'}(-l,-l')$ 
in general \cite{mh} and
that for symmetric systems (equilateral 3-disk-system with identical
disks, 2-disk with identical disks): $\Gamma_{jj'}^\ast (l,l') =
\Gamma_{j'j}(l,l')$ (see \cite{gr}). 
The right-hand side of eq.\equa{qm} is the
starting point for the semiclassical reduction, as every single 
term is
guaranteed to exist.
The properties of \equa{qm} can be summarized as follows:

\noindent 1. The product of the $n$ 1-disk determinants in \equa{qm} 
results from
the incoherent scattering where the $n$-disk problem 
is treated as $n$ 
single-disk problems.\\ 
2.
The whole expression \equa{qm} respects unitarity, since ${\bf
S}^{(1)}$ is unitary by itself 
[because of   $ \{H^{(2)}_{m} (z)\}^\ast =
H^{(1)}_{m}(z^\ast)$] and since the quotient on
the r.h.s.\ of \equa{qm} is manifestly unitary.\\
3.
The determinants on the r.h.s.\ in \equa{qm} run over the 
multi-index $L$.
This is the proper form to make the symmetry 
reductions in the multi-space,
e.g., for the equilateral 3-disk system 
(with disks of the same size) we have
\be
    \Det{ } {\bf M}_{\rm 3\mbox{-}disk} 
    = \det{      } {\bf M}_{A1}\, \det{     } {\bf M}_{A2} 
          \,\left ( \det{     } {\bf M}_{E} \right )^2 \ , 
 \label{3disk}
\ee
and    for the 2-disk system (with disks of the same size)
\be
    \Det{ } {\bf M}_{\rm 2\mbox{-}disk} = 
   \det{      } {\bf M}_{A1}\, \det{      } {\bf M}_{A2} 
     \, \det{      } {\bf M}_{B1} \,\det{      } {\bf M}_{B2} \; ,
 \label{2disk}
\ee  etc.
In general, if the disk configuration is characterized by a finite 
point symmetry group ${\cal G}$, we have
\be
  \Det{ } {\bf M}_{n{\rm\mbox{-}disk}}
 = \prod_r \left ( \det{   } {\bf M}_{D_r}(k) \right )^{d_r} \; ,
  \label{n_rep_disk} 
\ee
where the index $r$ runs over all 
conjugate classes of the symmetry group
${\cal G}$ and $D_r$ is the $r^{\,\rm th}$ 
representation of dimension 
$d_r$~\cite{mh}. [See \cite{Hammermesh} for notations and 
\cite{Lauritzen,cvi_eck_93} for the semiclassical
analog.]\  
A simple check that $\Det{}{\bf M}(k)$ has been split up 
correctly is the power of $\Ho{m}{a_j}$ Hankel functions 
(for fixed $m$ with 
$-\infty <m <+\infty$) appearing in the
denominator of  
$\prod_r \left ( \det{   } {\bf M}_{D_r}(k) \right )^{d_r}$ 
which has to be the same as in $\Det{} {\bf M}(k)$ 
which in turn has to
be the same as in 
$\prod_{j=1}^n \left (\det{} {\bf S}^{(1)}(ka_j)\right)$.
Note that on the l.h.s.\ 
the determinants are calculated in the multi-space $\{L\}$.
If the $n$-disk system is totally symmetric, i.e,  
none of the disks are special in size and position,
the reduced determinants on the r.h.s.\ are calculated 
in the normal (desymmetrized) space $\{l\}$, however, now 
with respect to the origin of the disk in
the fundamental domain and with ranges given by the corresponding
irreducible representations. If some of the $n$-disk 
are still special in size or position
(e.g., three equal disks in a row~\cite{wr96}), 
the determinants on the r.h.s.\
refer to a corresponding symmetry-reduced multi-space.
This is the symmetry reduction on the exact quantum-mechanical level.
The symmetry reduction 
can be most easily shown if one uses again the
trace-class properties of ${\bf M}-{\bf 1} \equiv {\bf A}$
\be
 \fl
   \Det{ } {\bf M} 
 = \exp\left \{ - \sum_{N=1}^\infty \frac{(-1)^N}{N} 
  \Tr{ }\left[ {\bf A}^N \right ] 
           \right \} 
 = \exp\left \{ - \sum_{N=1}^\infty \frac{(-1)^N}{N} \Tr{ }
      \left[{\bf U} {\bf A}^N {\bf U}^\dagger \right ] 
           \right \} \nno \\ 
 \lo= \exp\left \{ - \!\sum_{N=1}^\infty \frac{(-1)^N}{N} \Tr{ }
      \left[\left ( {\bf U} {\bf A} {\bf U}^\dagger \right )^N\right ] 
           \right \} 
 = \exp\left \{ -\! \sum_{N=1}^\infty \frac{(-1)^N}{N} \Tr{ }
      \left[{\bf A}^N_{\rm block}\right ] 
           \right \} \; , \nonumber
\ee
where ${\bf U}$ is unitary transformation which makes ${\bf A}$ 
block-diagonal in a
suitable basis spanned by the complete set $\{|m;j\rangle\}$.
These operations are allowed because of the trace-class-property of 
${\bf A}$ and the boundedness of the unitary matrix ${\bf U}$
(see the appendix).

As the right-hand side of eq.\equa{qm} splits into a product of
one-disk determinants and the ratio of two 
mutually complex conjugate
genuine $n$-disk determinants, which are all 
well defined individually, 
the semiclassical reduction can be performed for the one-disk and the
genuine multi-disk determinants separately. 
In \cite{aw_unpub} 
the  semiclassical expression for
the determinant of the 1-disk {\bf S}-matrix is 
constructed in analogous 
fashion to the semiclassical constructions of \cite{aw_chaos}:
\be
  \det{ } {\bf S}^{(1)}(ka) \approx  
               \left\{ {\rm e}^{-{\rm i} \pi N(ka)} \right\}^2 
        \, \frac
         { \left \{ \prod_{\ell=1}^{\infty} 
          \left[1 - {\rm e}^{-{\rm i} 
 2\pi {\bar \nu}_\ell(ka)}\right] \right\}^2 } 
         { \left \{ \prod_{\ell=1}^{\infty} 
          \left[1 - {\rm e}^{+{\rm i} 
       2\pi        \nu_\ell(ka)}\right] \right\}^2 }
 \label{sc1disk} 
\ee    
with the creeping term~\cite{franz,vwr_prl} 
\be
\fl
    \nu_\ell (ka) &=& ka + {\rm e}^{+{\rm i}\pi/3} 
             (ka/6)^{1/3}q_\ell+\cdots 
  = ka +{\rm i} \alpha_\ell (ka) +\cdots\; ,
  \label{nu_k} \\
\fl
  {\bar \nu}_\ell (ka) &=& ka + {\rm e}^{-{\rm i}\pi/3} 
           (ka/6)^{1/3}q_\ell +\cdots 
 = ka -{\rm i} ( \alpha_\ell (k^\ast a))^\ast +\cdots 
 = \left[\nu_\ell(k^\ast a)\right ]^\ast \; , 
  \label{bar_nu_k}
\ee
and $N(ka)= (\pi a^2 k^2)/4\pi + \cdots$ being
the leading term in the Weyl approximation for the 
staircase function of the 
wave number eigenvalues in the disk interior. {}From the 
point of view of
the scattering particle
the interior domains of the disks are excluded relatively to the
free evolution without scattering obstacles 
(see, e.g., \cite{scherer}),
hence the negative sign in front of the Weyl term. 
For the same reason
the subleading boundary term  has a Neumann structure, 
although the disks themselves obey Dirichlet boundary conditions. 
Let us abbreviate the r.h.s.\ of \equa{sc1disk} 
for a specified disk $j$ as
\be
 \det{ }{\bf S}^{(1)}(ka_j) \semiclass 
   \left \{ {\rm e}^{-{\rm i} \pi N(ka_j)} \right \}^2
      \, \frac{ { {\widetilde Z}_{\em l}^{(1)} (k^\ast a_j) }^\ast}
            {{\widetilde Z}_{\em l}^{(1)} (ka_j)}\, 
\frac{ { {\widetilde Z}_{\em r}^{(1)} (k^\ast a_j) }^\ast}
            {{\widetilde Z}_{\em r}^{(1)} (k a_j)}\ ,
\ee
where ${\widetilde Z}_{\em l}^{(1)}(ka_j)$ and $
{\widetilde Z}_{\em r}^{(1)}(k a_j)$
are the {\em diffractional} zeta functions (here and in the following
semiclassical zeta functions {\em with} diffractive corrections
shall be labelled
by a tilde)
for creeping
orbits around the $j$th disk  in the {\em left}-handed 
sense and the {\em right}-handed sense,
respectively.

The genuine multi-disk determinant
$\Det{} {\bf M}(k)$ (or $\det{} {\bf M}_{D_r}(k)$ in the case of
symmetric disk configurations) is organized according to the cumulant
expansion \equa{cumulant-sum} which, in fact, 
{\em is} the defining prescription for
the evaluation of the determinant of an 
infinite matrix under trace-class
property.
 Thus, the
cumulant arrangement is automatically imposed onto the semiclassical
reduction. Furthermore, the quantum-mechanical cumulants
satisfy
the Plemelj-Smithies recursion relation \equa{cumulant-recursion}
and can therefore solely be expressed  
by the quantum-mechanical traces $\Tr{}{\bf A}^m(k)$.
In ref.~\cite{aw_unpub} the 
semiclassical reduction of the traces, see eq.\equa{trace-semiclass},
has been derived. If this result is inserted back into
the Plemelj-Smithies recursion formula,   
the semiclassical equivalent of the 
exact cumulants arise. These are nothing but
the semiclassical curvatures \equa{curvature-recursion}, 
see~\cite{fredh,pinball,artuso}. Finally,
after the curvatures are 
summed up according to eq.\equa{curvature-sum},
it is clear that the 
the semiclassical reductions  of the  determinants  in \equa{qm} 
or \equa{n_rep_disk} are 
the Gutzwiller-Voros spectral determinants
(with creeping 
corrections) in the curvature-expansion-regularization. 
In the case where intervening disks ``block out'' 
ghost orbits~\cite{Berry_KKR,bb_2}), the corresponding orbits have
to be pruned, see \cite{aw_unpub}.
In summary, we have
\be
    \Det{ }{\bf M}(k) &\semiclass & 
 {\widetilde Z}_{GV} (k)|_{\rm curv.\,reg.} \; ,\\
    \det{ } {\bf M}_{D_r}(k) &\semiclass & 
{\widetilde Z}_{D_r} (k)|_{\rm curv.\,reg.} 
\ee   
where creeping corrections are included in 
the semiclassical zeta functions.   
The semiclassical limit of the r.h.s.\ of \equa{qm}
is  
\be
 \fl
  \det{ } {\bf S}^{(n)} (k) 
       =
\left\{ \prod_{j=1}^{n} 
 \det{ } {\bf S}^{(1)}(ka_j) \right \} \,
            \frac{\Det{ } { {\bf M}(k^\ast)}^\dagger}
{\Det{ } {\bf M}(k)} 
           \nno \\
 \lo{ \semiclass } 
       \left\{ \prod_{j=1}^n 
        \left ( {\rm e}^{-{\rm i} \pi N(ka_j )} \right )^{2}
      \, \frac{ {  {\widetilde Z}_{ {\em l} }^{(1)}
      (k^\ast a_j) }^\ast }
            {{\widetilde Z}_{ {\em l}}^{(1)} (k a_j)}\, 
      \, \frac{ {  {\widetilde Z}_{ {\em r}}^{(1)}
             (k^\ast a_j)  }^\ast }
            {{\widetilde Z}_{ {\em r}}^{(1)} (k a_j)}
        \, \right \}\,
            \frac{ { {\widetilde Z}_{GV}(k^\ast) }^\ast}
  {{\widetilde Z}_{GV}(k)}\; ,
    \label{gen}
\ee
where we now
suppress the qualifier 
$\cdots|_{\rm curv.\,reg.}$.
For systems which allow for complete symmetry 
reductions (i.e., equivalent disks with $a_j =a\ \forall j$.) the 
semiclassical reduction reads
\be
 \fl \det{ } {\bf S}^{(n)}(k) 
   =  
    \left\{  \det{ }{\bf S}^{(1)} (ka)\right \}^n 
     \frac{ \prod_r \left \{ \det{   } 
      { {\bf M}_{D_r} (k^\ast) }^\dagger
 \right \}^{d_r}}
 { \prod_r \left \{ \det{   } { {\bf M}_{D_r} (k) }
 \right \}^{d_r}}
   \nno \\
  \lo{\semiclass }
         \left \{ {\rm e}^{-{\rm i} \pi N(ka)} \right \}^{2n}
      \, \left \{
\frac{  {{\widetilde Z}_{\em l }^{(1)}(k^\ast a)}^\ast}
            {{{\widetilde Z}_{\em l}^{(1)} (ka)}}\, 
        \frac{{ {\widetilde Z}_{\em r}^{(1)}
              (k^\ast a)}^\ast}
            {{{\widetilde Z}_{\em r}^{(1)} (ka)}} \right \}^n
            \frac{ \prod_r \left \{{ 
          {\widetilde Z}_{D_r} (k^\ast)}^\ast 
          \right\}^{d_r}} 
              {\prod_r \left 
             \{ {{\widetilde Z}_{D_r}(k)} \right \}^{d_r}}
  \label{gen_sym_full}
\ee
in obvious correspondence. [See \cite{Lauritzen,cvi_eck_93} for
the symmetry reductions of the Gutzwiller-Voros zeta function.]
These equations do not
only give a relation between exact quantum mechanics and semiclassics
at the poles, 
but for {\em any} value of $k$ in the allowed $k$ region
(e.g., ${\rm Re}\, k > 0$).
There is  the caveat that the semiclassical limit and the
cumulant limit might not commute in general 
and that the curvature expansion
has a finite domain of convergence~\cite{fredh,fredh2,EckRuss}.

It should be noted that for {\em bound} systems the idea to focus 
not only on the positions of the zeros (eigenvalues) of the zeta
functions, but also on 
their analytic structure and their values taken
elsewhere was studied in refs.~\cite{KeatSieb94,Keat97}.

\section{Discussion\label{chap:end}}
\setcounter{equation}{0}
We have shown that \equa{qm} is a well-defined starting-point
for the investigation of the spectral properties of the 
exact quantum-mechanical scattering of a point particle from a
finite system of non-overlapping disks
in 2 dimensions. The genuine coherent multi-disk scattering decouples
from the incoherent superposition of $n$ single-disk problems. 
We furthermore demonstrated that 
\equa{gen} [or, for symmetry-reducible problems, 
equation \equa{gen_sym_full}] closes the gap between the
quantum mechanical and the semiclassical description 
of these problems.
Because the link involves determinants of infinite matrices with
trace-class kernels, the defining cumulant expansion
automatically induces
the curvature expansion for the semiclassical spectral function. 
We have also shown that in $n$-disk scattering systems unitarity
is preserved on the semiclassical level.

The result of \equa{gen} is compatible with Berry's 
expression for the integrated spectral density in Sinai's
billiard (a {\em bound} $n\to \infty$ disk system, 
see equation (6.11) of 
\cite{Berry_KKR}) and -- in general -- with the 
Krein-Friedel-Lloyd sums
\equa{friedel_sum}.
However, all the factors in the first line of 
the expressions \equa{gen}
and \equa{gen_sym_full} are not just of formal nature, but shown 
to be {\em finite}
except at the
zeros of the Hankel functions, $\Ho{m}{a}$ 
and $\Ht{m}{a}$, at the zeros of
the various determinants and on the negative real  $k$ axis,
since ${\bf M}(k)-{\bf 1}$ and 
${\bf S}^{(1)}(k)- {\bf 1}$ are ``trace-class'' 
almost everywhere in
the complex $k$-plane. 

The semiclassical
expressions (second lines of \equa{gen} and \equa{gen_sym_full}) 
are finite, if the zeta functions follow the induced 
curvature expansion and if the limit $m\to\infty$ exists 
also semiclassically [the curvature limit $m\to \infty$ 
and the semiclassical limit 
${\rm Re}\,k \to \infty$ do not have to commute].
The curvature regularization  is the semiclassical
analog to the well-defined 
quantum-mechanical cumulant expansion.
This justifies the formal manipulations 
of \cite{scherer,pinball,moroz}. 

Furthermore, even semiclassically, unitarity is automatically 
preserved in scattering problems (without
any reliance on re-summation techniques 
\`{a} la Berry and Keating\cite{berry_keats} 
which are necessary and only
applicable in
bound systems), 
since
\be 
  \det{ } {{\bf S}^{(n)} (k)}^\dagger 
         = \frac{1}{\det{ } {\bf S}^{(n)}(k^\ast)}
\ee 
is valid both quantum-mechanically (see the first lines of \equa{gen} 
and   \equa{gen_sym_full}) and 
semiclassically (see
the second lines of \equa{gen} and \equa{gen_sym_full}). 
There is the caveat that the curvature-regulated 
semiclassical zeta function 
has a finite domain of convergence defined
by the poles of the dynamical zeta function in the lower complex 
$k$-plane~\cite{fredh,fredh2,EckRuss}.
Below this boundary line the semiclassical zeta function has to be 
truncated at finite order in the curvature expansion~\cite{cvw96}.
Thus, under the stated conditions unitarity is preserved for $n$-disk
scattering systems on the
semiclassical level. 
On the other hand, unitarity can therefore not be used
in scattering problems to gain any constraints on the 
structure of
${\widetilde Z}_{GV}$ as it could in bound systems,
see \cite{berry_keats}.

To each (quantum-mechanical or semiclassical) pole of
$\det{ } {\bf S}^{(n)}(k)$ in the lower complex $k$-plane determined
by a zero of $\Det{}{\bf M}(k)$ there 
belongs a zero of $\det{ } {\bf S}(k)$
in the upper complex $k$-plane (determined by a zero of 
$\Det{}{\bf M}(k^\ast)$ with the same ${\rm Re}\, k$ value, 
but opposite ${\rm Im}\, k$.
We have also demonstrated that the zeta functions 
of the pure 1-disk scattering and 
the genuine multi-disk scattering decouple,
i.e.,
the 1-disk poles do not influence the position of 
the {\em genuine} multi-disk poles.
However, $\Det{ }{\bf M}(k)$ does not only possess zeros, 
but also poles. The
latter exactly cancel the poles of the product of the 1-disk
determinants, $\prod_{j=1}^{n} \det{ }{\bf S}^{(1)}(ka_j)$, 
since both involve the same ``number'' and ``power'' 
of $\Ho{m}{a_j}$ Hankel
functions in the denominator. The same is true for the poles of 
${\Det{ } { {\bf M}(k^\ast)}^\dagger}$ and the {\em zeros} of 
 $\prod_{j=1}^{n} \det{ } {\bf S}^{(1)}(ka_j)$, 
since in this case the ``number'' of $\Ht{m}{a_j}$
Hankel functions  in the denominator of the former and the
numerator of the latter is the same --- see also Berry's 
discussion on the same cancellation 
in the integrated spectral density
of Sinai's billiard, equation (6.10) of \cite{Berry_KKR}. 
Semiclassically, this cancellation corresponds to a 
removal of the additional
creeping contributions of topological length zero, 
$1/(1-\exp({\rm i} 2\pi \nu_{\ell}))$,
from ${\widetilde Z}_{GV}$ by the 1-disk diffractive
zeta functions, ${\widetilde Z}_{\em l}^{(1)}$ and $
{\widetilde Z}_{\em r}^{(1)}$. 
The orbits of topological length zero result
from the geometrical sums over additional creepings 
around the single disks,
$\sum_{n_{w}=0}^{\infty} 
(\,\exp({\rm i} 2\pi \nu_\ell)\,)^{n_w}$ 
(see \cite{vwr_prl}). They
multiply the
ordinary creeping paths of non-zero topological length.
Their cancellation
is very important in situations
where the disks nearly touch, since in such geometries
the full circulations of 
creeping orbits around any of
the touching disks should clearly be suppressed, as it now is.
Therefore, it is important to keep consistent account of
the diffractive contributions in the semiclassical reduction.
Because of the decoupling of the one-disk from the 
multi-disk determinants,
a direct clear comparison of the quantum mechanical 
cluster phase shifts of
$\Det{} {\bf M}(k)$ with the semiclassical ones 
of the Gutzwiller-Voros
zeta function $Z_{GV}(k)$ is possible, which 
otherwise would be only small
modulations on the dominating single-disk phase shifts (see 
\cite{cvw96,aw_unpub}).

In the standard cumulant expansion [see \equa{cumulant-sum} 
with the 
Plemelj-Smithies recursion formula \equa{cumulant-recursion}] as well 
as in the
curvature expansion [see \equa{curvature-sum} with 
\equa{curvature-recursion}] there
are large cancellations 
involved  which become more and more dramatic the
higher the cumulant order is. 
Let us order -- without loss of generality  --
the eigenvalues of the trace-class operator ${\bf A}$ as 
follows:
\be
 |\lambda_1|\geq |\lambda_2|\geq \cdots \geq |\lambda_{i-1}|\geq 
|\lambda_i |\geq 
|\lambda_{i+1}|\geq 
 \cdots \; .\nn
\ee
This is always possible because the 
sum over the moduli of the eigenvalues
is finite for trace-class operators.
Then, in the standard (Plemelj-Smithies) cumulant evaluation of 
the determinant there are cancellations of
big numbers, e.g., at the $l^{\,\rm th}$ cumulant order ($l>3)$, 
all the intrinsically
large ``numbers''
 $\lambda_1^l$, $\lambda_1^{l-1}\lambda_2$, $\dots$, 
$\lambda_1^{l-2}\lambda_2\lambda_3$, $\dots$  
and many more have to cancel
out, such that the r.h.s.\ of 
\be
 \det{ } ( {\bf} 1 + z {\bf A}) = \sum_{l=0}^{\infty}\, z^l \,
 \sum_{j_1 < \cdots < j_l}
 \lambda_{j_1}({\bf  A}) \cdots \lambda_{j_l} ({\bf A} )  .
  \label{det_cum_red_new}
\ee
is finally left over.  
Algebraically, the large cancellations in
the exact quantum-mechanical calculation do not matter of course.  
However, if
the determinant is calculated numerically, 
large   cancellations might spoil
the result or even the convergence. 
Moreover, if further approximations
are made as, e.g., the transition 
from the exact cumulant to the semiclassical
curvature expansion, these large cancellations might 
be potentially dangerous.
Under such circumstances the underlying (algebraic) absolute
convergence of the 
quantum-mechanical cumulant expansion cannot simply
induce the convergence of the 
semiclassical curvature expansion, since
{\em large} semiclassical ``errors'' can completely change the
convergence properties.  

In summary, 
the non-overlapping disconnected $n$-disk systems have the
great virtue that -- although classically completely
hyperbolic and for some systems even chaotic -- they are
quantum-mechanically
{\em and} semiclassically ``self-regulating'' and also
``self-unitarizing'' and still simple enough
that the semiclassics can be
studied directly, 
independently of the Gutzwiller formalism, and then
compared with the latter.

\section*{Acknowledgements}

A.W.\ would like to thank the 
Niels-Bohr-Institute and Nordita for hospitality, 
and especially Predrag
Cvitanovi\'{c}, Per Rosenqvist, Gregor Tanner, Debabrata Biswas
and Niall Whelan for many discussions.  
M.H.\ would like to thank Friedrich Beck for fruitful discussions
and helpful advice.  

\begin{appendix}

\section{Existence of the n-disk {\bf S}--matrix 
and its determinant\label{app:suppl}}
\setcounter{equation}{0}
Gaspard and Rice \cite{gr} derived in a formal way an expression
for the {\bf S}--matrix for the 3-disk repeller. 
We used the same techniques
in order to generalize 
this result to repellers consisting of $n$ disks
of {\em different} radii \cite{aw_unpub,mh},
\be
   {\bf S}^{(n)} = {\bf 1} - {\rm i}\,{\bf T} \; , \qquad
   {\bf T} = {\bf B}^j \cdot {\bf D}^j \; 
        \label{S_n_rep}
\ee
\be     \label{CAM_2}
  {\bf C}^j = {\bf B}^{j'} \cdot {\bf M}^{j' j}
\ee
\be
                 \label{SCMD_2} 
   {\bf S}^{(n)}  =  {\bf 1} - {\rm i}\, 
         {\bf C}^{j} \cdot ({\bf M}^{-1})^{jj'} 
                                  \cdot {\bf D}^{j'} \; .
 \label{s_summary}
\ee
${\bf S}^{(n)}$ denotes the {\bf S}--matrix 
for the $n$-disk repeller
and ${\bf B}^{j}$ parametrizes the gradient 
of the wavefunction on 
the boundary of the disk $j$. 
The matrices {\bf C} and {\bf D} describe
the coupling of the incoming and 
outgoing scattering waves, respectively,
to the disk $j$ and the matrix {\bf M} is the genuine multi-disk
``scattering'' matrix with eliminated single-disk properties.
{\bf C}, {\bf D} and {\bf M} are given by eqs.~\equa{Cmatrix},
\equa{Dmatrix} and \equa{Mmatrix} respectively.
The derivations of the expression for  
{\bf S}--matrix~\equa{SCMD_2} 
and 
of its determinant (see section~\ref{chap:link}) 
are of purely formal character as all the matrices involved are
of infinite size. Here, we will show that the performed
operations are all well-defined. For this purpose, the trace-class 
(${\cal J}_1$)
and Hilbert--Schmidt (${\cal J}_2$)   operators will play a 
central role. 

\subsection*{Trace class and determinants of infinite matrices}
We will briefly summarize  the definitions and most important
properties for trace-class
and Hilbert-Schmidt matrices and operators and for determinants over
infinite dimensional matrices, 
refs.~\cite{rs1,bs_adv,bs,gohberg,kato} 
should be consulted for details and proofs.

\noindent An operator ${\bf A}$ 
is called {\bf trace class}, ${\bf A} \in 
{\cal J}_1$, if and only if, for every 
orthonormal basis, $\{\phi_n\}$:
\be
  \sum_n |\langle \phi_n, {\bf A} \phi_n \rangle| < \infty \;.
\ee
An operator ${\bf A}$ is called {\bf Hilbert-Schmidt},  ${\bf A} \in 
{\cal J}_2$, if and only if, 
for every orthonormal basis, $\{\phi_n\}$:
\be
  \sum_n \| {\bf A} \phi_n \|^2 < \infty \;.
\ee
\noindent The most important properties
of the trace and Hilbert-Schmidt classes
can be summarized as (see \cite{rs1,bs}):
(a) ${\cal J}_1$ and ${\cal J}_2$ are $\ast$ideals., i.e., they
are vector spaces closed under scalar 
multiplication, sums, adjoints, and
multiplication with bounded operators.
(b) ${\bf A}\in {\cal J}_1$ if and only if ${\bf A}={\bf  BC}$ with
${\bf B},{\bf C}\in {\cal J}_2$.
(c) For any operator ${\bf A}$, we have ${\bf A}\in {\cal J}_2$ if
$\sum_n \| {\bf A} \phi_n \|^2 < \infty$ for a single basis.
(d) For any operator ${\bf A}\geq 0$,
we have ${\bf A}\in {\cal J}_1$ if
$\sum_n |\langle \phi_n, {\bf A} \phi_n \rangle| < \infty$ 
for a single basis.

Let ${\bf A} \in  {\cal J}_1$, then the determinant $\det
({\bf 1} + z {\bf A})$ exists \cite{rs1,bs_adv,bs,gohberg,kato},
it is an entire and analytic function of $z$ 
and it can be expressed by
the {\em Plemelj-Smithies formula:}\ Define $\alpha_m({\bf A})$
for ${\bf A}\in {\cal J}_1$ by
\be
 \det{ } ( {\bf 1}+ z{\bf A}) = \sum_{m=0}^{\infty} z^m
 \frac{\alpha_m ({\bf A})}{m!} \ .
 \label{ps_formula}
\ee
Then $\alpha_m({\bf A})$ is given by the $m\times m$ determinant
\be
 \alpha_m({\bf A}) = \left | \begin{array}{ccccc}
    \Tr{ }({\bf A}) & m-1     &  0  & \cdots & 0 \\
  \Tr{ }({\bf A}^2) & \Tr{ }({\bf A})    &  m-2 & \cdots & 0 \\
   \Tr{ }({\bf A}^3) &  \Tr{ }({\bf A}^2) &  \Tr{ }({\bf A})
    & \cdots & 0 \\
  \vdots & \vdots & \vdots & \vdots & \vdots \\
  \Tr{ }({\bf A}^m) &   \Tr{ }({\bf A}^{(m-1)}) &
    \Tr{ }({\bf A}^{(m-2)}) & \cdots & \Tr{ }({\bf A})
 \end{array} \right |
\ee
with the understanding that $\alpha_0({\bf A})\equiv 1$ and
$\alpha_1({\bf A})\equiv \Tr{ } ({\bf A})$.
Thus the cumulants $c_m({\bf A}) \equiv \alpha_m({\bf A})/ m!$
(with  $c_0 ({\bf A}) \equiv  1 $)
satisfy
the recursion relation
\be
 c_m({\bf A}) &=& \frac{1}{m}\sum_{k=1}^{m} 
(-1)^{k+1} c_{m-k}({\bf A})\,
   \Tr{ }({\bf A}^k) \qquad {\rm for}\ m \geq 1 \; . \nonumber
\ee
The most important properties of these determinants are:
(i) If ${\bf A},{\bf B} \in {\cal J}_1$, then
$ \det{ } ({\bf 1} + {\bf A}) \det{ } ({\bf 1}+ {\bf B} )
 = \det{ }\left
 ( {\bf 1} +{\bf A}+{\bf B} + {\bf A B} \right ) 
 = \det{ }\left[ ({\bf 1} +{\bf A})({\bf 1} +{\bf B}) \right ]
 = \det{ }\left[ ({\bf 1} +{\bf B})({\bf 1} +{\bf A}) \right ]$.
(ii) If ${\bf A}\in {\cal J}_1$ and ${\bf U}$ unitary, then
$ \det{ } \left ({\bf U}^{\dagger} 
( {\bf 1}+{\bf A} ) {\bf U} \right )
= \det{ } \left ({\bf 1}+{\bf U}^{\dagger}{\bf A} {\bf U} \right )
 = \det{ } ({\bf 1} + {\bf A})$.
(iii) If ${\bf A}\in {\cal J}_1$, then
$({\bf 1}+ {\bf A})$ is invertible 
if and only if $\det{ }({\bf 1}+{\bf A})
\neq 0$.
(d) For any ${\bf A} \in {\cal J}_1$,
\be
\det{ } ( {\bf 1}+  {\bf A} ) 
 = \prod_{j=1}^{ N( {\bf A} ) } \left[ 1
+ \lambda_j ( {\bf A} ) \right ] \; ,
  \label{det_prod}
\ee
where here and in the following
$ \{ \lambda_j ( {\bf A} ) \}_{j=1}^{ N( {\bf A} )}$ 
are the eigenvalues of
${\bf A}$ counted with algebraic multiplicity ($N({\bf A})$ can of
course be infinite).

\vspace{0.3cm}

Now we can return to the actual problem.
The ${\bf S}^{(n)}$--matrix is given by~\equa{S_n_rep}.
The {\bf T}-matrix is  trace-class on the positive
real $k$ axis ($k>0$), 
as it is the product 
of the matrices ${\bf D}^j$ and ${\bf B}^j$ which
will turn out to be trace-class or, respectively,
bounded there
(see \cite{rs1,bs_adv} for the definitions). 
Again formally,  we have used  that
${\bf C}^j={\bf B}^{j'}{\bf M}^{j'j}$ 
implies the relation ${\bf B}^{j'}=
{\bf C}^{j}({\bf M}^{-1})^{jj'}$. 
Thus, the existence of ${\bf
M}^{-1}(k)$ has to be shown, too -- except at isolated poles in the 
lower complex $k$ plane 
below the real $k$ axis and on the branch cut on the negative real 
$k$ axis which
results
from the branch cut of the defining Hankel functions. 
As we will prove later, 
${\bf M}(k)-{\bf 1}$ is trace-class, except of course 
at the above mentioned 
points in the $k$ plane.  
Therefore, using property (iii),
we only have to show that 
$\Det{}{\bf M}(k)\neq 0$ in order to guarantee the
existence of ${\bf M}^{-1}(k)$.  At the same time, 
${\bf M}^{-1}(k)$ will be proven to be
bounded 
as all its eigenvalues and the product 
of its eigenvalues are then finite.
The existence of these eigenvalues  
follows from the trace-class property of
${\bf M}(k)$ which together with $\Det{}{\bf M}(k)\neq 0$ 
guarantees the finiteness of the eigenvalues and their 
product~\cite{rs1,bs_adv}.

We have normalized ${\bf M}$ in such a way
that we simply have ${\bf B}={\bf C}$ 
for the scattering from a single disk. Note that
the structure of the matrix ${\bf C}^j$ does not dependent
on the fact whether the point particle 
scatters only from a single disk or
from $n$ disks. 
The functional form \equa{Cmatrix}
shows that ${\bf C}$ cannot have poles 
on the real positive $k$ axis ($k>0$) 
in agreement
with the structure of the 
${\bf S}^{(1)}$--matrix [see equation 
\equa{S1disk}]. 
If the origin of the coordinate system 
is put into the origin of the disk,
the matrix  ${\bf S}^{(1)}$ is 
diagonal. 
In the same basis ${\bf C}$ becomes diagonal. 
One can easily see that
${\bf C}$  
has no zero eigenvalue on
the positive real $k$ axis and that it will be trace-class.
So neither ${\bf C}$ nor the 1--disk (or for that purpose 
the $n$--disk ) {\bf S} matrix can 
possess poles or zeros on the real positive
$k$ axis. The statement about ${\bf S}^{(n)}$ 
follows simply from the unitarity of the {\bf S}-matrix which can 
be checked easily.
The fact that 
$|\det{} {\bf S}^{(n)}(k)|=1$ 
on the positive real $k$ axis cannot be used
to disprove that $\Det{} {\bf M}(k)$ could be zero there 
[see equation \equa{qm}]. However, if  
$\Det{} {\bf M}(k)$ were zero there, 
this ``would-be'' pole must cancel
out of  ${\bf S}^{(n)}(k)$. 
Looking at formula \equa{s_summary}, this pole
has to cancel out against a zero from 
${\bf C}$ or ${\bf D}$ where both
matrices are already fixed on the 1-disk level.  
Now, it follows from \equa{det_prod} that 
${\bf M}(k)$ (provided that
${\bf M}-{\bf 1}$ has been proven  trace-class)  has only one chance 
to make trouble
on the positive real $k$ axis, 
namely, if at least one of its eigenvalues 
(whose existence is guaranteed) 
becomes zero. On the other hand ${\bf M}$ has still to satisfy 
${\bf C}^j = {\bf B}^{j'}{\bf  M}^{j'j}$.
Comparing the left and the right-hand side of
$
 |{\bf C}^{\ \ j}_{mm}(k)| = 
          | {\bf B}^{\ \ j'}_{ml} {\bf M}_{lm}^{j' j}|  
$
in the eigenbasis of {\bf M}
and having in mind 
that ${\bf C}^j(k)$ cannot have zero eigenvalue for $k>0$
one finds a contradiction 
if the corresponding eigenvalue
of ${\bf M}(k)$ were zero. 
Hence ${\bf M}(k)$ is
invertible on the real positive $k$ axis, 
provided, as mentioned now
several times, ${\bf M}(k)- {\bf 1}$ is trace-class.
{}From the existence of the inverse relation 
${\bf B}^{j'}={\bf C}^{j} ({\bf M}^{-1})^{jj'}$ and the to be shown 
trace-class property
of ${\bf C}^{j}$ and the 
boundedness of $({\bf M}^{-1})^{jj'}$ follows the
boundedness of ${\bf B}^{j}$ and 
therefore the trace-class property of
the $n$-disk {\bf T}-matrix, ${\bf T}^{(n)}(k)$, 
except at the above excluded
$k$-values. 

What is left for us to do is to prove
\begin{description}
\item[\ (a)\ ] ${\bf M}(k)-{\bf 1}\in {\cal J}_1$ 
for all $k$, except at
the poles of $\Ho{m}{a_j}$ and for $k\leq 0$,
\item[\ (b)\ ] ${\bf C}^j(k), {\bf D}^j(k) \in {\cal J}_1$ 
with the exception
of the same $k$-values  mentioned in {\bf (a)},
\item[\ (c)\ ] ${\bf T}^{(1)}(ka_j) \in {\cal J}_1$ (again
with the same exceptions
as in {\bf (a)}) where ${\bf T}^{(1)}$ is the {\bf T}-matrix
of the 1--disk problem,
\item[\ (d)\ ] ${\bf M}^{-1}(k)$ does not only exist, but is bounded.
\end{description} 
Under these conditions all the manipulations of section~\ref{chap:link} 
[equations \equa{recoupl} and \equa{qm}] are
justified and ${\bf S}^{(n)}$, as in \equa{Smatrix}, and
$\det{} {\bf S}^{(n)}$, as in \equa{qm}, are shown to exist.

\subsection*{Proof of ${\bf T}^{(1)}(ka_j))\in {\cal J}_1$  }
The  {\bf S}--Matrix for the $j^{\,\rm th}$ disk is given by 
\be     \label{S_1disk}
   {\bf S}^{(1)}_{ml}(ka_j) 
 = -\frac{H^{(2)}_{l}(ka_j)}{H^{(1)}_{l}(ka_j)}
         \delta_{ml} \; .  
\ee
Thus ${\bf V}\mbox{$\equiv$}-{\rm i}\,{\bf T}^{(1)}(ka_j) =
{\bf S}^{(1)}(ka_j)\mbox{$-$}{\bf 1}$ is diagonal.
Hence, we can write ${\bf V} = {\bf U} {\bf | V|}$
where ${\bf U}$ is diagonal and unitary, 
and therefore bounded. What is
left to show is that 
${\bf |V|}\in {\cal J}_1$
We just
have to show in a special orthonormal basis 
(the eigenbasis) that 
\be
\sum_{l=-\infty}^{+\infty} {\bf |V|}_{ll} 
 = \sum_{l=-\infty}^{+\infty}
  2 \left |\frac{J_l(ka_j)}{H^{(1)}_{l}(ka_j)}\right | < \infty \; ,
\ee
since ${\bf |V|}\geq 0$ by definition (see property (d)).
The convergence of this series can be shown easily using the asymptotic
formulae for Bessel and Hankel functions for large orders, $\nu \to \infty$,
$\nu$ real:
\be
 J_\nu (ka) 
  \sim \frac{1}{\sqrt{2\pi\nu}} \left(\frac{e ka}{2\nu}\right)^\nu
 \; , \quad  
H^{(1)}_\nu (ka) \sim -i\sqrt{\frac{2}{\pi\nu}} 
   \left(\frac{e ka}{2\nu}\right)^{-\nu}
 \label{Bess_asymp}
\ee
(see e.g. 
\cite{Abramowitz}).
{}From this equation  follows the mathematical justification for the
impact parameter (or angular momentum) 
truncation in the semiclassical resolution of
the {\em single} disks, $|m| \le \frac{e}{2} ka$.
This limit should not be confused with the truncation in the
curvature order resulting from the finite resolution of the
repelling set of the $n$-disk problem, see ref.~\cite{cvw96}.
Under these asymptotic formulae and the summation of the resulting
geometrical series, the trace-class property of   
${\bf |V|}\in {\cal J}_1$ and
${\bf S}^{(1)}-{\bf 1}\in {\cal J}_1$ follows immediately. 
That in turn  means that  $\det{} {\bf S}^{(1)}(ka_j)$ exists 
and  also that the product 
$\prod_{j=1}^n \det{} {\bf S}^{(1)}(ka_j) < \infty$ if 
$n$ is finite (see \cite{rs1,bs_adv}).
Note that 
the limit ${n\to \infty}$ does not exist in general.

\subsection*{ Proof of ${\bf A}(k) \equiv {\bf M}(k) - {\bf 1} \in 
{\cal J}_1$}
The determinant of the characteristic matrix ${\bf M}(k)$ 
is defined, if
${\bf A}(k) \in {\cal J}_1$. In order to show this, we split ${\bf A}$ 
into
the product of two operators which -- as we will show -- 
are both Hilbert-Schmidt. Then 
the product is trace-class (see property (b)).

Let therefore ${\bf A} = {\bf E} \cdot {\bf F}$ with
{\bf A}= {\bf M} $-$ {\bf 1} as given in  \equa{Mmatrix}.
In order to simplify the 
decomposition of ${\bf A}$, we choose one of the
factors, namely, 
${\bf F}$,  as a diagonal matrix. 
Let therefore 
\be  \label{def_F2}
  {\bf F}^{jj'}_{ll'}  = 
  \frac{\sqrt{H^{(1)}_{2l}(k\alpha a_j)}}{H^{(1)}_l(ka_j)}\,
 \delta^{jj'}\,\delta_{ll'} 
   \;, \qquad \alpha >2 \; .
\ee
This ansatz already excludes 
the zeros of the  Hankel functions
${H^{(1)}_l(ka_j)}$  
and also the  negative real $k$ axis (the branch cut
of the Hankel functions for  $k\leq 0$) 
from our final proof of ${\bf A}(k)\in {\cal J}_1$. 
First, we have to show that 
$\|{\bf F}\|^2=\sum_{j}\sum_{l} ({\bf F^\dagger
F})^{jj}_{ll} < \infty$. We start with 
\beq
  \| {\bf F} \|^2 
   \le  \sum_{j=1}^n 2 \sum_{l=0}^\infty
    \frac{|H^{(1)}_{2l}(k\alpha a_j)|}
     {|H^{(1)}_l(ka_j)|^2} \; \equiv \;
    \sum_{j=1}^n 2 \sum_{l=0}^\infty {\rm a}_l \; .
\eeq
This expression restricts our proof to $n$-disk configurations with 
$n$ {\em finite}. 
Using the asymptotic expressions 
for the Bessel and Hankel 
functions of large orders  \equa{Bess_asymp} 
(see e.g.\ 
\cite{Abramowitz}), it is easy to prove 
the absolute convergence of $\sum_l {\rm a}_l$ 
in the case $\alpha > 2$.
Therefore
$\|{\bf F}\|^2 < \infty$ and because of property (c) 
we get ${\bf F} \in {\cal J}_2$.

\vspace{0.3cm}

We now investigate the second factor {\bf E}. 
We have to show the convergence of 
\beq
 \fl
  \|{\bf E}\|^2 = \sum_{j,j'=1 \atop j \neq j'}^n
    \left( \frac{a_j}{a_{j'}} \right)^2 \sum_{l,l'=-\infty}^\infty
          {\rm a}_{ll'}  \; , \qquad
      {\rm a}_{ll'}  =
    \frac{|J_l(ka_j)|^2 |H^{(1)}_{l-l'}(kR_{jj'})|^2}
         {|H^{(1)}_{2l'}(k\alpha a_{j'})|}  
\eeq
in order to prove that also ${\bf E}\in {\cal J}_2$.
Using the same techniques as 
before the convergence of $\sum_l {\rm a}_{ll'}$
for $(1+\epsilon) a_j < R_{jj'}$, $\epsilon>0$, 
as well as the convergence of
$\sum_{l'} {\rm a}_{ll'}$ for 
$\alpha a_{j'} < 2 R_{jj'}$, $\alpha > 2$,
can be shown.
We must of course show the convergence  of 
$\sum_{l,l'} {\rm a}_{ll'}$ 
for 
the case 
$l,l' \to \infty$ as well. Under the asymptotic behavior of the
Bessel and Hankel functions of large order \equa{Bess_asymp},
it is easy to see that it suffices to prove the convergence of
$\sum_{l,l'=0}^\infty b_{ll'}$, where
\be
 b_{ll'} = \frac{(l+l')^{2(l+l')}}
   { l^{2l} {l'}^{2l'}}
   \left( \frac{a_j}{R_{jj'}} \right)^{2l}
   \left( \frac{\alpha}{2} \frac{a_{j'}}{R_{jj'}} \right)^{2l'} \; .
\ee
In order to show the convergence
of the double sum, we introduce new summation indices
($M,\, m$), namely $2M:=l+l'$ and $m:=l-l'$.
Using first Stirling's formula 
for large powers $M$ and then applying the
binomial formula in order to perform the 
summation over $m$, the convergence
of $\sum_{l,l'=0}^\infty b_{ll'}$ can be shown, provided that 
$a_j + \frac{\alpha}{2} a_{j'} < R_{jj'}$.
Under this condition
the  operator {\bf E} belongs to the class  of Hilbert--Schmidt
operators~(${\cal J}_2$).
 \newline \hspace{2cm}
In summary, this means:
${\bf E(k) \cdot F(k) = A}(k) \in {\cal J}_1$ for those $n$ 
disk configurations for which the number of disks is finite and the 
disks neither overlap nor touch and for
those values of $k$ which 
lie neither on the zeros of the Hankel functions
$\Ho{m}{a_j}$ nor on the negative real $k$ axis ($k\leq 0$).
The zeros of the Hankel functions 
$H^{(2)}_m(k^\ast a_j )$ are then automatically excluded, too.
The zeros of the
Hankel functions $\Ho{m}{\alpha a_j}$ 
in the definition of ${\bf E}$ are
cancelled by the corresponding zeros of 
the same Hankel functions in
the definition of ${\bf F}$ and can therefore be removed, 
i.e., a slight
change in $\alpha$ readjusts the positions of the zeros 
in the complex 
$k$ plane such that they can 
always be moved  to non-dangerous places.

\subsection*{Proof of ${\bf C}^j, {\bf D}^j \in {\cal J}_1$}
The expressions for ${\bf D}^j$ and
${\bf C}^j$ 
can be found in \equa{Dmatrix} and \equa{Cmatrix}. 
Both matrices contain -- for a fixed value of $j$ --  
only the information of the single disk scattering.
As in the proof of ${\bf T}^{(1)}\in {\cal J}_1$, 
we go to the eigenbasis
of ${\bf S}^{(1)}$. In that basis both matrices 
${\bf D}^j$ and ${\bf C}^j$ become diagonal.
Using the same techniques as in the 
proof of  ${\bf T}^{(1)}\in {\cal J}_1$,
we can show that
${\bf C}^j$ and  ${\bf D}^j$ are trace-class. 
In summary, we have ${\bf D}^j \in  {\cal J}_1$ for all $k$
as the Bessel functions  which define that matrix 
possess neither poles nor branch cuts. 
The matrix 
${\bf C}^j$ is traceclass for almost every $k$, except at the
zeros of the Hankel functions $\Ho{m}{a_j}$ 
and the branch cut of these
Hankel functions on the negative real $k$ axis ($k\leq 0$).

\subsection*{Existence and boundedness of ${\bf M}^{-1}(k)$}
$\Det{} {\bf M}(k)$ exists almost everywhere,
since ${\bf M}(k)-{\bf 1}\in {\cal J}_1$, except at the zeros of
$\Ho{m}{a_j}$ and on the negative real $k$ axis ($k\leq 0$).
Modulo these points ${\bf M}(k)$ is analytic. 
Hence, the points of the complex 
$k$ plane  with $\Det{} {\bf M}(k) =0$ are isolated.
Thus almost
everywhere ${\bf M}(k)$ can be 
diagonalized and the product of the eigenvalues
weighted by their degeneracies is finite and nonzero.
Hence, where $\Det{} {\bf M}(k)$ is defined and nonzero,
${\bf M}^{-1}(k)$ exists, it
can be diagonalized and the product of its eigenvalues
is finite. 
In summary, ${\bf M}^{-1}(k)$ is bounded and 
$\Det{}{\bf M}^{-1}(k)$ exists
almost everywhere in
the complex $k$ plane. 
\end{appendix}

\end{document}